\documentclass[12pt,preprint]{aastex}
\usepackage[dvips]{color}
\usepackage{multicol}
\usepackage{epsf}
\usepackage{epsfig} 
\usepackage{amssymb}
\newcommand{\bm}[1]{{\mbox{\boldmath$#1$}}}
\shorttitle{Relativistic VLBI Experiment}
\shortauthors{Kopeikin Sergei and Fomalont Ed}
\begin{document}
\title{The Measurement of the Light Deflection from
Jupiter: \\Theoretical Interpretation}
\author{S. M. Kopeikin}
\affil{Department of Physics and Astronomy, University of
Missouri-Columbia, MO 65211, USA}
\email{kopeikins@missouri.edu}

\begin{abstract}\noindent
Equations of light, propagating from quasar to observer on earth, are integrated in the time-dependent gravitational field of the solar system by making use of either retarded or advanced solutions of the Einstein field equations. This technique allows to separate explicitly the effects associated with the propagation of gravity from those associated with light in the integral expression for the relativistic time delay of light. We prove that the relativistic correction to the Shapiro time delay discovered by Kopeikin (ApJ, 556, L1, 2001) changes sign if one retains direction of the light propagation but replaces retarded with advanced solution of the gravitational field equations. Hence, this correction is due to the propagation of gravity and allows to measure its speed. Relativistic VLBI experiment conducted by Fomalont and Kopeikin in September 2002 proves that the propagation of gravitational field is characterized by the retarded potentials.  
\end{abstract}
\newpage 

\section{Introduction}
\noindent 
Einstein's equations of general relativity have a hyperbolic character \citep{ll, wald}. From mathematical point of view a hyperbolic wave equation allows in general two types of solutions described respectively by retarded and advanced potentials \citep{synge,ah}. It is often supposed that the advanced solution is unphysical because it violates the physical principle of causality. However, the theory of general relativity does not exclude the advanced solution per se. Making a choice between the retarded and advanced solutions is a matter of a boundary condition imposed on the Einstein equations. Although the causality-preserving boundary condition is reasonable from theoretical point of view, one does not know a priori what kind of condition is realized for the gravitational field in real world. For this reason, experimental study of the nature of the boundary condition for Einstein's equations has a great importance for foundations of theoretical physics. Timing observations of binary pulsar PSR 1913+16 indicates that the retarded solution of the Einstein equations is the most appropriate for the description of orbital dynamics of the neutron stars \citep{wt}. Nevertheless, an independent experimental verification of the retarded nature of the gravity propagation is required. 

In this paper we solve null geodesic equations for light propagating from a distant source (quasar) in time-dependent gravitational field of moving bodies (solar system) by making use of the mathematical technique invented by \cite{k1997} and further elaborated in \citep{ks1999, kb2002}. We derive the relativistic delay in the time of propagation of light and analyze how it relates to the retarded and advanced solutions of the Einstein equations. This give us an unambiguous criterion for distinguishing the relativistic effects associated with the propagation of gravity from those related to the propagation of light -- the question which was a matter of confusion for some researchers \citep{a,will}.  

In this paper we assume that the propagation of light from quasar towards observer always goes along the electromagnetic null-cone from the past to the future. On the other hand, we allow the wave solutions of the Einstein equations describing propagation of gravitational field to be both retarded and advanced. Difference between them is incorporated to the time variable $s_\epsilon$ that (for each solution) is an implicit function of the coordinate time $t$ given by the gravity null-cone equation 
\begin{equation}
\label{1}
s_\epsilon=t+\frac{\epsilon}{c}|{\bm x}-{\bm x}_a(s_\epsilon)|\;,
\end{equation}    
where ${\bm x}$ is a field point, ${\bm x}_a(s_\epsilon)$ is position of $a$th gravitating body \footnote{Notice that the variable $s_\epsilon$ is different for each body as it depends on the body's position ${\bm x}_a$ as well.}  ($a=1,2,...,N$) depending on the time argument $s_\epsilon$. Numerical parameter $\epsilon\equiv -c/c_g$, where $c$ is the limiting speed of special relativity and $c_g$ is the 'speed-of-gravity' parameter taking either positive or negative values such that in general relativity $|c_g|=c$. 

In order to distinguish the gravity-propagation effects from those associated with light we shall operate separately with the advanced gravity null cone characterized formally by the condition $-c_g=c$, and with the retarded gravity null cone singled out by the condition $c_g=c$. Thus, the parameter $\epsilon$ in Eq. (\ref{1}) takes two different numerical values 
\begin{equation}
\label{1a}
\epsilon=\left\{\begin{array}{r@{\quad:\quad}l}-1&\mbox{retarded time\,,}\\
+1&\mbox{advanced time\,,}\end{array}\right.
\end{equation}
such that $\epsilon^2=1$. 
The parameter $\epsilon$ plays a supplementary role helping to track down the presence of the gravity propagation effects in the solutions of both Einstein's and the light-ray-propagation equations. This tracking property of the parameter $\epsilon$ is used for providing unique and unambiguous physical interpretation of the relativistic effect in the VLBI time delay predicted by \cite{k2001} and observed by \cite{fk2003}. 

In what follows we rely upon the system of units such that $G=c=1$. Boldface italic letters denote spatial vectors, e.g, ${\bm a}\equiv a^i=(a^1,a^2,a^3)$. The dot between two spatial vectors is the Euclidean scalar product, ${\bm a}{\bm\cdot} {\bm b}=a^1b^1+a^2b^2+a^3b^3$. The cross between two spatial vectors denotes the Euclidean vector product, ${\bm a}\times{\bm b}\equiv \varepsilon_{ijk}a^jb^k$, where $\varepsilon_{ijk}$ is the Levi-Civita symbol such that $\varepsilon_{123}=+1$. Roman indices run from 1 to 3 while Greek indices run from 0 to 3. Repeated indices assume the Einstein summation rule. Greek indices are raised and lowered with the Minkowski metric $\eta_{\alpha\beta}={\rm diag}(-1,+1,+1,+1)$. 

\section{The Retarded and Advanced Solutions of the Einstein Equations}
\noindent
We assume that the gravitational field is weak everywhere so that the metric tensor $g_{\alpha\beta}(t,{\bm x})$$=\eta_{\alpha\beta}+h_{\alpha\beta}(t,{\bm x})$, where $h_{\alpha\beta}(t,{\bm x})$ is the perturbation of the gravitational field, and $t$ and ${\bm x}$ are time and spatial coordinates. 

Let gravitational field be generated by the stress-energy tensor of moving point-like massive particles \citep{ll}
\begin{eqnarray}
\label{2} 
T^{\alpha\beta}(t, {\bm x})&=&\sum_{a=1}^N
M_a \gimel_a^{-1}(t)\,u_a^\alpha(t)\,
u_a^\beta(t)\,
\delta\bigl({\bm x}-{\bm
x}_a(t)\bigr)\;,
\end{eqnarray}
where $a=1,2,...,N$
enumerates gravitating bodies of the solar system, $M_a$
is the (constant) rest mass of the $a$th body,
${\bm x}_a(t)$ are time-dependent spatial coordinates of the $a$th body, 
${\bm v}_a(t)= d{\bm x}_a(t)/dt$ is the orbital
velocity of the $a$th body, $u_a^\alpha=\gimel_a (1,\,{\bm v}_a)$ is the
four-velocity of the $a$th body, $\gimel_a=\bigl(1-v_a^2\bigr)^{-1/2}$,
and $\delta({\bm x})$ is a 3-dimensional Dirac's delta-function.

We shall work in the harmonic coordinates \citep{wei} with the origin at the barycenter of the solar system. Formal mathematical solution of the Einstein equations in the harmonic coordinates and in the linearized post-Minkowskian approximation \citep{bel,d} is given by the {\it Li\'enard-Wiechert} tensor potential
\begin{eqnarray}
\label{3} 
h^{\alpha\beta}_\epsilon(t,{\bm x})&=& 4\sum_{a=1}^N\;\frac{M_a}{\gimel_a(s_\epsilon)}\;\frac{ u_{a}^\alpha(s_\epsilon) u_{a}^\beta(s_\epsilon)+(1/2)\eta^{\alpha\beta}}
{r_a(s_\epsilon)+\epsilon{\bm v}_a(s_\epsilon)\cdot {\bm r}_a(s_\epsilon)}\;,
\end{eqnarray}
where ${\bm r}_a(s_\epsilon)={\bm x}-{\bm x}_a(s_\epsilon)$, $r_a(s_\epsilon)=|{\bm r}_a(s_\epsilon)|$, ${\bm v}_a(s_\epsilon)=d{\bm x}_a(s_\epsilon)/ds_\epsilon$, and the time argument $s_\epsilon=s_\epsilon(t,{\bm x})$ is determined as an implicit
solution of the retarded ($\epsilon=-1$) or advanced ($\epsilon=+1$) gravity null-cone Eq. (\ref{1}).
At this stage it is obvious that because we were solving the Einstein equations only gravity propagation with the speed $c$ is involved in Eq. (\ref{3}) through the retarded (advanced) time argument $s_\epsilon$.

Non-linear corrections to the linear solution (\ref{3}) of the Einstein equations can be obtained by making use of the iterative post-Minkowskian approximations \citep{bel,d}. This non-linear metric tensor should be substituted to the equations of motion of the bodies comprising the  gravitating system under consideration. Solution of the equations of motion gives us the parametric trajectory ${\bm x}_a(t)$ of each body orbiting around the barycenter of the $N$-body system with the instantaneous velocity ${\bm v}_a(t)=d{\bm x}_a(t)/dt$, and the acceleration ${\bm a}_a(t)=d{\bm v}_a(t)/dt$. We shall assume that such solution of the body's equations of motion has been found and used in the expression for the metric tensor perturbation given by Eq. (\ref{3}). More involved discussion of this problem in the framework of the post-Minkowskian approximations is given by \cite{bel} and \cite{d}. 

\section{Light-Ray Trajectory Parameterization}
\noindent
We consider motion of light in the non-stationary gravitational 
field described by the metric $g_{\alpha\beta}$ and assume that light has no back action on the gravitational field. Hence, one can use 
equations of light geodesics. 
Let the motion of light be defined by the
initial-boundary conditions 
\begin{equation}
{\bm x}(t_{0})={\bm x}_{0}\;, \hspace{2 cm} 
{\displaystyle {d{\bm x}(-\infty ) \over dt}}
={\bm k}\;,
\label{7}
\end{equation}
where ${\bm k}^{2}=1$. These conditions define the coordinates 
${\bm x}_{0}$ of the photon at the instant of emission of light, $t_{0}$, and direction of its 
velocity at the past null infinity \citep{wald}. 

The original null geodesic equations of light propagation are rather 
complicated. These equations can be simplified and
recast to the integrable form \citep{ks1999, kb2002}. In the case of weak gravitational field the integration is performed by making use of 
iterations \citep{k1997}. One starts from the unperturbed trajectory of the 
light ray that is a straight line
\begin{eqnarray}
\label{8}
x^i(t)&=&x^i_N(t)=x^i_0+k^i\left(t-t_0\right)\;,
\end{eqnarray}
where the coordinate time $t$ is a running parameter along the light ray, and $t_0$, $x^i_0$, $k^i={\bm k}$ have been defined in Eq. (\ref{7}). 
In this approximation velocity of the photon is 
$\dot{x}^i=k^i$, and is considered as constant in the next iteration involving the metric tensor perturbation $h_{\alpha\beta}$.

It is convenient to introduce a new independent parameter $\sigma$ along the
photon's trajectory according to the rule 
\begin{eqnarray}
\label{9}
\sigma&=& {\bm k}\cdot{\bm x}_N(t)=t-t_0+{\bm k}\cdot{\bm x}_{0}\;.
\end{eqnarray}
The time $t_0$ of the light emission 
corresponds to the
numerical value of the parameter $\sigma_0={\bm k}\cdot{\bm x}_{0}$, and
the numerical value of the parameter $\sigma=0$ corresponds to the instant  
\begin{eqnarray}
\label{uuu}
t^{\ast}&=&
t_0-{\bm k}\cdot{\bm x}_{0}\;,
\end{eqnarray} 
that is the time of the closest approach of the unperturbed trajectory of 
light ray to the origin of the coordinate system.
We emphasize that the numerical value of the moment $t^{\ast}$ is constant for
a chosen trajectory of light ray and
depends only on the space-time coordinates of the point of emission
of the photon and direction of its propagation. Thus, we find the relationship
\begin{eqnarray}
\label{10}
t&\equiv&t^{\ast}+\sigma\;,
\end{eqnarray}
which reveals that the variable $\sigma$ is negative from the point of emission 
up to the point of the closest approach, $x^i(t^\ast)={\xi}^i$, to the origin of the coordinate system, and is 
positive 
otherwise. The differential identity $dt=d\sigma$ is valid and, for this 
reason, 
the integration along the light ray's path with respect to time $t$ can be 
replaced by the integration with respect to variable $\sigma$.

Making use of the parameter $\sigma$, the equation of the unperturbed trajectory 
of 
the light ray can be represented as
\begin{eqnarray}
\label{11}
x^i(\sigma)&=&x^i_N(\sigma)=k^i \sigma+\xi^i\;,
\end{eqnarray}
and the distance, $r(\sigma)=|{\bm x}_N(t)|$, of the photon from the origin of 
the coordinate system reads
\begin{eqnarray}
\label{12}
r(\sigma)&=&\sqrt{\sigma^2+d^2}\;.
\end{eqnarray}
The constant vector $\xi^i={\bm{\xi}}={\bm k}\times ({\bm x}_{0}
\times {\bm k})={\bm k}\times \left({\bm x}_N(t)\times {\bm k}\right)$ 
is called the impact parameter of the unperturbed trajectory of
the light ray with respect to the origin of the coordinate system, $d=|{\bm{\xi}}|$ is the (Euclidean) length of the impact parameter. 
We notice that the vector ${\bm{\xi}}$ is 
transverse
to the vector ${\bm k}$ in the Euclidean sense, that is ${\bm k}{\bm\cdot}{\bm\xi}=0$, where "${\bm\cdot}$" denotes the Euclidean dot product. Vector $\xi^i$ 
is directed from the origin of the coordinate system towards the point of the 
closest approach of the unperturbed path of the light ray to this origin. This
vector plays auxiliary role in our discussion and has no
essential physical meaning as it can be changed by the shift of
the origin of the coordinates. Time of the closest approach $t^{\ast}$ is a convenient parameter being useful in the following calculations but it does not enter the final formula for the relativistic time delay as it should be expected on the general physical ground.

\section{Relativistic Time Delay}
\noindent
Time of propagation of light from the point $x_0^i$ to $x^i_1\equiv x^i(t_1)$ is \citep{k2003}
\begin{equation}
\label{qer}
t_1-t_0=|{\bm x}_1-{\bm x}_0|+\Delta(t_1,t_0)\;,
\end{equation}
where 
\begin{equation}
\label{ty1}
\Delta(t_1,t_0)={1\over 2}\,k^\alpha k^\beta \int_{t_0}^{t_1}h_{\alpha\beta}\left(t,{\bm x}_N(t)\right)\,dt \;,
\end{equation}
$t_1$ is the time of observation, and the (Minkowskian) null vector of the light propagation $k^\alpha=(1, k^i)$ is directed from the past to the future.
For the sake of simplicity we shall work out the subsequent calculations only up to the linear terms with respect to the (non-constant) velocity $v_a$ of the moving bodies. This is presently enough for the purposes of the relativistic astrometry in the solar system.

Substitution of Eq. (\ref{3}) to (\ref{ty1}) and replacement of the variable $t\rightarrow\sigma$, yields 
\begin{equation}
\label{ty}
\Delta(t_1,t_0)=2\sum_{a=1}^N M_a\int_{\sigma_0}^{\sigma_1}\frac{\left(1-{\bm k}\cdot{\bm v}_a(s_\epsilon)\right)^2}{r_a(s_\epsilon)+\epsilon {\bm v}_a(s_\epsilon)\cdot {\bm r}_a(s_\epsilon)}\,d\sigma \;,
\end{equation}
where ${\bm r}_a(s_\epsilon)={\bm {\xi}}+{\bm k}\sigma-{\bm 
x}_a(s_\epsilon)$, and the variable $s_\epsilon$ relates to the coordinate time $\sigma$ via Eq. (\ref{1}) of the gravity null cone taken on the unperturbed light ray trajectory (electromagnetic null cone) at the point at which the light particle is at the time $\sigma$
\begin{equation}
\label{17}
\sigma+t^{\ast}=s_\epsilon-\epsilon|{\bm {\xi}}+{\bm k}\sigma-{\bm 
x}_a(s_\epsilon)|\;.
\end{equation}
We remind once again that the parameter $\epsilon=\pm 1$ indicates the presence of the gravity-propagation effects. 

Differentiation of Eq. (\ref{17})
yields an exact relationship between the 
differentials of the time variables $\sigma$ and $s_\epsilon$ 
\begin{equation}
\label{newvar}
\frac{d\sigma}{r_a+\epsilon{\bm v}_a\cdot{\bm r}_a}=\frac{ds_\epsilon}{r_a+\epsilon{\bm k}\cdot{\bm
r}_a}\;,
\end{equation}
where ${\bm r}_a={\bm x}_N(t)-{\bm x}_a(s_\epsilon)$, and the denominator in the left side of this equation coincides exactly with the denominator of the metric tensor perturbation in Eq. (\ref{3}). Transformation (\ref{newvar}) allows to replace integration with respect to coordinate time $\sigma$ to that with respect to the retarded (advanced) time $s_\epsilon$ which is simpler and more informative.

In order to take integral (\ref{ty}) one introduces a new variable $y\equiv r_a+\epsilon{\bm k}\cdot{\bm r}_a$ and calculates it on the gravity null cone Eq. (\ref{17}).
\begin{equation}
\label{23}
y_\epsilon\equiv r_a+\epsilon{\bm k}\cdot{\bm r}_a=-\epsilon \left(t^{\ast}+{\bm k}\cdot{\bm
x}_a(s_\epsilon)-s_\epsilon\right)\;.
\end{equation} 
We get
\begin{eqnarray}
\label{new}
dy_\epsilon&=&\epsilon \left(1-{\bm k}\cdot {\bm v}_a(s_\epsilon)\right)ds_\epsilon\;,
\end{eqnarray}
so that the above integral (\ref{ty}) is recast to 
\begin{eqnarray}
\label{45}
\Delta(t_1,t_0)=2\epsilon \sum_{a=1}^N M_a\int_{y_{\epsilon 0}}^{y_{\epsilon 1}}\Bigl[1-{\bm k}\cdot{\bm v}_a\left(s_\epsilon(y_\epsilon)\right)\Bigr]d\ln y_\epsilon \;.
\end{eqnarray}
Integration by parts yields
\begin{eqnarray}
\label{er1}
\Delta(t_1,t_0)&=&
2\epsilon \sum_{a=1}^N M_a\Biggl[\Bigl(1-{\bm k}\cdot{\bm v}_a(s_{\epsilon 1})\Bigr)\ln y_{\epsilon 1} -\Bigl(1-{\bm k}\cdot{\bm v}_a(s_{\epsilon 0})\Bigr)\ln y_{\epsilon 0}\Biggr]\;,
\end{eqnarray}
where
\begin{eqnarray}
\label{sd}
y_{\epsilon 1}&=&r_{1a}+\epsilon{\bm k}\cdot{\bm r}_{1a}\;,\\
y_{\epsilon 0}&=&r_{0a}+\epsilon{\bm k}\cdot{\bm r}_{0a}\;,
\end{eqnarray}
with ${\bm r}_{1a}={\bm x}_1-{\bm x}_a(s_{\epsilon 1})$, ${\bm r}_{0a}={\bm x}_0-{\bm x}_a(s_{\epsilon 0})$, and the retarded (advanced) times $s_{\epsilon 1}$ and $s_{\epsilon 0}$ are connected to the time of observation, $t_1$, and emission of light, $t_0$, via the gravity null cone Eq. (\ref{1}). Eq. (\ref{er1}) does not assume that the velocities of the gravitating bodies are constant. It only means that we have neglected accelerations when performing the integral.  

\section{Differential VLBI time delay}
\noindent
Let us assume now that there are two earth-based VLBI stations and
the front of an electromagnetic wave from quasar propagates towards the earth. Quasars are almost at the edge of the visible universe and they do not reveal any annual parallax and/or proper motion \citep{16}. For the array of the VLBI stations spread out across the earth one can assume that the wave front of quasar's electromagnetic radiation is plane and characterized by a single wave vector $k^\alpha=(1,{\bm k})$. Taking two rays from the wave front, directed from the quasar towards the first and second VLBI stations,
and subtracting Eq. (\ref{er1}) for the
first light ray from that for the second ray yields
\begin{equation}
\label{6}
t_2-t_1=-{\bm K}{\bm\cdot}{\bm B}+\Delta(t_1,t_2)\;,
\end{equation}
where $\Delta(t_1,t_2)\equiv\Delta(t_2,t_0)-\Delta(t_1,t_0)$, ${\bm K}=-{\bm k}$ is the unit vector from the
barycenter of the solar system to the quasar (quasar's annual parallax is neglected), and ${\bm B}={\bm x}_2(t_1)-{\bm x}_1(t_1)$ is the barycentric
baseline between the two VLBI stations. Here
$t_1$ and $t_2$ are the barycentric coordinate times of arrival of the
electromagnetic signal from the quasar to the first and second VLBI stations respectively, and ${\bm
x}_1$ and ${\bm x}_2$ are spatial coordinates of the first and second
VLBI stations with respect to the barycentric frame of the solar
system that is chosen as the primary non-rotating reference frame
\citep{16}. We omitted in Eq. (\ref{6}) the terms which are proportional to the velocity of motion of the second station as these terms are not crucial in discussion pertained to the interpretation of the relativistic VLBI experiment conducted on September 8, 2002 \citep{fk2003}. Further theoretical details of the relativistic algorithm of transformation of the barycentric to geocentric coordinates are contained in \citep{iers, 16}. They are not so important in the discussion that follows, but are essential for understanding how VLBI works in general. 

The difference $\Delta(t_1,t_2)$ 
is obtained from Eqs. (\ref{er1}), (\ref{6}) after long and tedious
calculations accompanied
by analysis of the residual terms similar to that done in
\citep{k2001}.  We have proved that with accuracy better than 1 ps,
\begin{eqnarray}
\label{7a} \Delta(t_1,t_2)&=&-2\epsilon\sum_{a=1}^N M_a
\left[1+{\bm K}{\bm\cdot}{\bm v}_a(s_{\epsilon 1})\right]\;
\ln\frac{r_{1a}(s_{\epsilon 1})-\epsilon{\bm K}{\bm\cdot}{\bm r}_{1a}(s_{\epsilon 1})}
{r_{2a}(s_{\epsilon 2})-\epsilon{\bm K}{\bm\cdot}{\bm r}_{2a}(s_{\epsilon 2})}\;,
\end{eqnarray}
where  ${\bm v}_a(s_{\epsilon 1})$ is
the velocity of the $a$th gravitating body at the retarded (advanced) time $s_{\epsilon 1}$, $r_{1a}=|{\bm r}_{1a}|$,
$r_{2a}=|{\bm r}_{2a}|$, ${\bm r}_{1a}(s_{\epsilon 1})={\bm
x}_{1}(t_1)-{\bm x}_{a}(s_{\epsilon 1})$ and ${\bm r}_{2a}(s_{\epsilon 2})={\bm
x}_{2}(t_2)-{\bm x}_{a}(s_{\epsilon 2})\;$; moreover, the retarded (advanced) times $s_{\epsilon 1}$ and
$s_{\epsilon 2}$ are calculated according to Eq. (\ref{1}) 
\begin{eqnarray}
\label{8a}
s_{\epsilon 1}&=&t_1+\epsilon |{\bm x}_1(t_1)-{\bm x}_a(s_{\epsilon 1})|\;,\\\label{8b}
s_{\epsilon 2}&=&t_2+\epsilon |{\bm x}_2(t_2)-{\bm x}_a(s_{\epsilon 2})|\;,
\end{eqnarray}
where again we notice the presence of the gravity-propagation parameter $\epsilon=\pm 1$. Thus, Eqs. (\ref{8a}), (\ref{8b}) reveal the true physical reason for the coordinates ${\bm x}_a$ of the bodies in Eq. (\ref{7a}) to be displaced from their present positions - the finite speed of propagation of gravity. 

Measuring relativistic time delay of radio signal propagating from quasar through the gravitational field of moving body allows to evaluate position of the body in the sky at the time of observation and compare it with that one would observe if position of the body were measured directly in optics by receiving light emitted by the body itself \citep{seid}. In the case of the retarded solution of the Einstein equations ($\epsilon=-1$) position of the body measured directly in optics and that measured indirectly through the realtivistic time delay coincide precisely, that is both optical and gravitational 'images' of the body experience the same aberration in complete agreement with general relativity. It is worth noting that for slowly-moving particles this effect can not be observed in the approximation under discussion because of a different structure of the perturbing gravitational force leading to the apparent cancellation of retarded effects in the solutions of the post-Newtonian equations of motion of the particles \citep{carlip}. This cancellation does not happen for light and the retardation of gravity can be observed experimentally by making use of high-precision VLBI technique \citep{fk2003}. 

\section{Retarded versus Advanced Solutions of the Einstein Equations in the VLBI Time Delay}
\noindent
Let us elaborate Eq. (\ref{7a}) for two different cases corresponding to the retarded $(\epsilon = -1)$ and advanced $(\epsilon = +1)$ solutions of the Einstein equations. This will help us to split the gravity-propagation effects from those associate with the propagation of light. The gravity propagation effects depend on the parameter $\epsilon$ and changes sign when one uses retarded solution instead of advanced one and vice versa.

\subsection{Retarded Solution}
\noindent
The retarded case of $\epsilon = -1$ has been already treated in the papers \citep{k2001, k2003} from various points of view. The effect of retardation of gravity appears in Eq. (\ref{7a})
as a displacement of the light-ray deflecting bodies from their
present ${\bm x}_a(t_i)$ to retarded positions ${\bm x}_a(s_{-i})$ $(i=1,2)$. We notice that velocities of the
gravitating bodies, ${\bm v}_a$, are small with respect to the fundamental 
constant $c$. Furthermore, the time taken by light ray to cross the solar system
is much smaller than 
orbital periods of Sun and other planets around the barycenter of the solar system. Thus, one can expand the retarded positions
${\bm x}_a(s_{-i})$ of the bodies in the Taylor series
around their present positions ${\bm x}_a(t_i)$ taken at the arrival
times $t_i$ to the first or second VLBI stations. Moreover, we shall assume that only the impact parameter $d_J$ of the light ray with respect to Jupiter is small. Hence \citep{k2001},
\begin{equation}
\label{imp} 
{\bm N}_{1J}\equiv{\bm r}_{1J}/r_{1J}=-\left(1-\frac{\theta^2}{2}\right){\bm
K}+\theta\;{\bm{n}}+O(\theta^3)\;,
\end{equation}
where hereafter the subscript $J$ refers to Jupiter, ${\bm n}$ is the unit vector of the impact parameter of the light ray with respect to Jupiter, and $\theta$ is a small angle in the sky between the directions towards the
undisturbed geometric position of the quasar and that of Jupiter.

After expanding body's retarded position ${\bm x}_a(s_{-i})$ around the instant $t_i$,
neglecting all terms quadratic with respect
to velocities of the gravitating bodies, proportional to
their accelerations and the products of body's velocities with the angle $\theta$, one gets
\begin{equation}
\label{9a} \Delta(t_1,t_2)=2\sum_{a=1}^N \frac{GM_a}{c^3}
\left(1+\frac{{\bm K}{\bm\cdot}{\bm v}_a}{c}
\right) \ln\frac{R_{1a}+{\bm K}{\bm\cdot}{\bm R}_{1a}}
{R_{2a}+{\bm K}{\bm\cdot}{\bm R}_{2a}}
-\frac{2GM_J}{c^4} \frac{({\bm v}_J{\times}{\bm K}){\bm\cdot}({\bm B}{\times}{\bm
K})}{R_{1J}+{\bm
K}{\bm\cdot}{\bm R}_{1J}}\;,
\end{equation}
where we have restored the constants $G$ and $c$ for the sake of convenience, ${\bm R}_{ia}={\bm
x}_{i}(t_i)-{\bm x}_{a}(t_i)$ is the difference between coordinates
of the $i$th VLBI station and $a$th gravitating body taken at
the time of arrival of radio signal to the $i$th station, ${\bm
v}_a\equiv{\bm v}_a(t_1)$, and all quantities in the second term
on the right side of Eq. (\ref{9a}) are also evaluated at
time $t_1$.

The second term in Eq. (\ref{9a}) was obtained as a consequence of expansion of the retarded time $s_-$ ($\epsilon=-1$) in the gravity null cone Eqs. (\ref{8a}), (\ref{8b}) which describe for this value of $\epsilon$ the effect of retardation of gravity. We emphasize that light from the quasar does not propagate along the gravity null cone that connects the body (Jupiter, Sun, etc.) and the light particle. Light propagates along the null world line, connecting the quasar and observer, which is not a part of the gravity null cone \footnote{This might be a reason for the misconceptions presented in \citep{a,will} and discussed by \cite{k2003} in more detail.}. Consequently, the origin of the second term in Eq. (\ref{9a}) is due to the propagation of gravity as it was shown by \cite{k2001}. 

To confirm our interpretation from different prospect we shall study the differential VLBI time delay for the case of the advanced solution of the Einstein equations. If the second term in the right side of Eq. (\ref{9a}) is due to the propagation of gravity, it must change sign in the advanced solution. This indeed takes place.

\subsection{Advanced Solution}
\noindent
Let us now take the advanced solution of Einstein equations, that is $\epsilon=+1$, and calculate the impact of the advanced gravitational field on the VLBI time delay. We shall first have from Eq. (\ref{7a})
\begin{eqnarray}
\label{ad1} \Delta(t_1,t_2)&=&-2\sum_{a=1}^N M_a
\Bigl(1+{\bm K}{\bm\cdot}{\bm v}_a(s_{+1})\Bigr)\;
\ln\frac{r_{1a}(s_{+1})-{\bm K}{\bm\cdot}{\bm r}_{1a}(s_{+1})}
{r_{2a}(s_{+2})-{\bm K}{\bm\cdot}{\bm r}_{2a}(s_{+2})}\;,
\end{eqnarray} 
where the advanced times $s_{+1}$ and $s_{+2}$ are computed from Eqs. (\ref{8a}), (\ref{8b}) with $\epsilon=+1$. We notice that the logarithmic function in Eq. (\ref{ad1}) can be recast as follows
\begin{eqnarray}
\label{ad2}
-\ln\frac{r_{1a}(s_{+1})-{\bm K}{\bm\cdot}{\bm r}_{1a}(s_{+1})}
{r_{2a}(s_{+2})-{\bm K}{\bm\cdot}{\bm r}_{2a}(s_{+2})}=\ln\frac{r_{1a}(s_{+1})+{\bm K}{\bm\cdot}{\bm r}_{1a}(s_{+1})}
{r_{2a}(s_{+2})+{\bm K}{\bm\cdot}{\bm r}_{2a}(s_{+2})}-2\ln\frac{|{\bm K}\times{\bm r}_{1a}(s_{+1})|}{|{\bm K}\times{\bm r}_{2a}(s_{+2})|}\;.
\end{eqnarray}
By making use of the Taylor expansion of ${\bm x}_a(s_{+2})$ around the instant $s_{+1}$ one can get 
\begin{eqnarray}
\label{ad3}\frac{|{\bm K}\times{\bm r}_{1a}(s_{+1})|}
{|{\bm K}\times{\bm r}_{2a}(s_{+2})|}&=&1+\frac{({\bm K}\times{\bm r}_{1a}){\bm\cdot}({\bm K}\times{\bm v}_a)}{|{\bm K}\times{\bm r}_{1a}(s_{+1})|^2}\Bigl(s_{+2}-s_{+1}\Bigr)+O(v_a^2)\;, \end{eqnarray}
where
\begin{eqnarray}
\label{ad4}
s_{+2}-s_{+1}&=&-{\bm K}{\bm\cdot}{\bm B}+{\bm N}_{1a}{\bm\cdot}{\bm B}+O(v_a)\;.
\end{eqnarray}

The second term in the right side of Eq. (\ref{ad3}) is negligibly small for any body of the solar system but might be large enough to be measured in the field of Jupiter. For Jupiter, one has
\begin{eqnarray}
\label{ad5}
-2\ln\frac{|{\bm K}\times{\bm r}_{1J}(s_{+1})|}{|{\bm K}\times{\bm r}_{2J}(s_{+2})|}&=& \frac{4({\bm K}{\bm\cdot}{\bm B})({\bm v}_J{\bm\cdot}{\bm n})}{r_{1J}\theta}\;,
\end{eqnarray} 
where we put together two terms in Eq. (\ref{ad4}) after making use of Eq. (\ref{imp}). One can see from Eq. (\ref{ad5}) that the second term in the right side of Eq. (\ref{ad2}) is by the factor $\theta$ smaller than the second term in the right side of Eq. (\ref{9a}). During the time of the VLBI experiment \citep{fk2003} $\theta\sim 3.7$ arcminutes and the second term in the right side of Eq. (\ref{9a}) was about 51 $\mu$as. Hence, the magnitude of the second term in the right side of Eq.(\ref{ad2}) is much less than 1 $\mu$as and can be neglected because the precision of the VLBI experiment on September 8, 2002 was $\sim$ 10 $\mu$as \citep{fk2003}.

Thus, we obtain that in the case of the advanced solution of the Einstein equations the differential VLBI time delay in Eq. (\ref{ad1}) is given by
\begin{eqnarray}
\label{ad6} \Delta(t_1,t_2)&=&2\sum_{a=1}^N M_a
\Bigl(1+{\bm K}{\bm\cdot}{\bm v}_a(s_{+1})\Bigr)\;
\ln\frac{r_{1a}(s_{+1})+{\bm K}{\bm\cdot}{\bm r}_{1a}(s_{+1})}
{r_{2a}(s_{+2})+{\bm K}{\bm\cdot}{\bm r}_{2a}(s_{+2})}\;,
\end{eqnarray}  
where $s_{+1}$ and $s_{+2}$ are advanced times related to the times of arrival $t_1$ and $t_2$ of radio signal to the first and second VLBI stations through the solutions of the advanced gravity cone Eqs. (\ref{8a}), (\ref{8b}) with $\epsilon=+1$. 

Taylor expansion of positions ${\bm x}_a(s_{+i})$ of the bodies around the coordinate times of arrival $t_i$ transforms Eq. (\ref{ad6}) to 
\begin{equation}
\label{ad7} \Delta(t_1,t_2)=2\sum_{a=1}^N {GM_a\over
c^3}\left(1+{{\bm K}{\bm\cdot}{\bm v}_a\over
c}\right) \ln\frac{R_{1a}+{\bm K}{\bm\cdot}{\bm R}_{1a}}
{R_{2a}+{\bm K}{\bm\cdot}{\bm R}_{2a}}
+\frac{2GM_J}{c^4} \frac{({\bm v}_J{\times}{\bm K}){\bm\cdot}({\bm B}{\times}{\bm
K})}{(R_{1J}+{\bm
K}{\bm\cdot}{\bm R}_{1J})}\;.
\end{equation}
Comparison of Eq. (\ref{ad7}) with Eq. (\ref{9a}) elucidates that the signs of the second terms in their right sides are opposite. Hence, these terms originates due to the effect of propagation of gravity and can be used to measure its speed. This interpretation has been given by \cite{k2001} and is confirmed by calculations given in the present paper. Interpretations given by \cite{a} and \cite{will} are erroneous and originates from misleading ideas.

\section{Experimental Verification of the Retarded Nature of the Einstein Equations}
\noindent
Let us recast Eqs. (\ref{9a}) and (\ref{ad7}) to the following form (c.f. Eq. (12) from \citep{k2001}) 
\begin{equation}
\label{ad8} \Delta(t_1,t_2)=2\sum_{a=1}^N {GM_a\over
c^3}\left(1+{{\bm K}{\bm\cdot}{\bm v}_a\over
c}\right) \ln\frac{R_{1a}+{\bm K}{\bm\cdot}{\bm R}_{1a}}
{R_{2a}+{\bm K}{\bm\cdot}{\bm R}_{2a}}
-(1+\delta)\,\frac{2GM_J}{c^4} \frac{({\bm v}_J{\times}{\bm K}){\bm\cdot}({\bm B}{\times}{\bm
K})}{(R_{1J}+{\bm
K}{\bm\cdot}{\bm R}_{1J})}\;,
\end{equation}
where the parameter $\delta=-(\epsilon+1)\equiv c/c_g-1$ with the speed of gravity $c_g=c$, if we operate with the retarded ($\epsilon=-1$) branch of the gravity null cone Eq. (\ref{1}), and $-c_g=c$, if the advanced ($\epsilon=+1$)  branch of the gravity null cone Eq. (\ref{1}) is used. Hence, $\delta=0$, if solutions of the Einstein gravitational field equations have a retarded nature, and $\delta=-2$, if the nature of the solutions of the Einstein equations is advanced. Furthermore, one could use a half-retarded plus half-advanced solutions of the Einstein equations in order to calculate the differential VLBI time delay. We have fulfilled such calculation and proved that it corresponds to the case of $c_g=\infty$ in Eq. (\ref{ad8}), that is $\delta=-1$ ($\epsilon=0$). Formally, it is equivalent to the case of the Newtonian theory in which the propagation of gravity is instantaneous with infinite speed \citep{k2003}.  

\begin{figure}
\noindent
\plotone{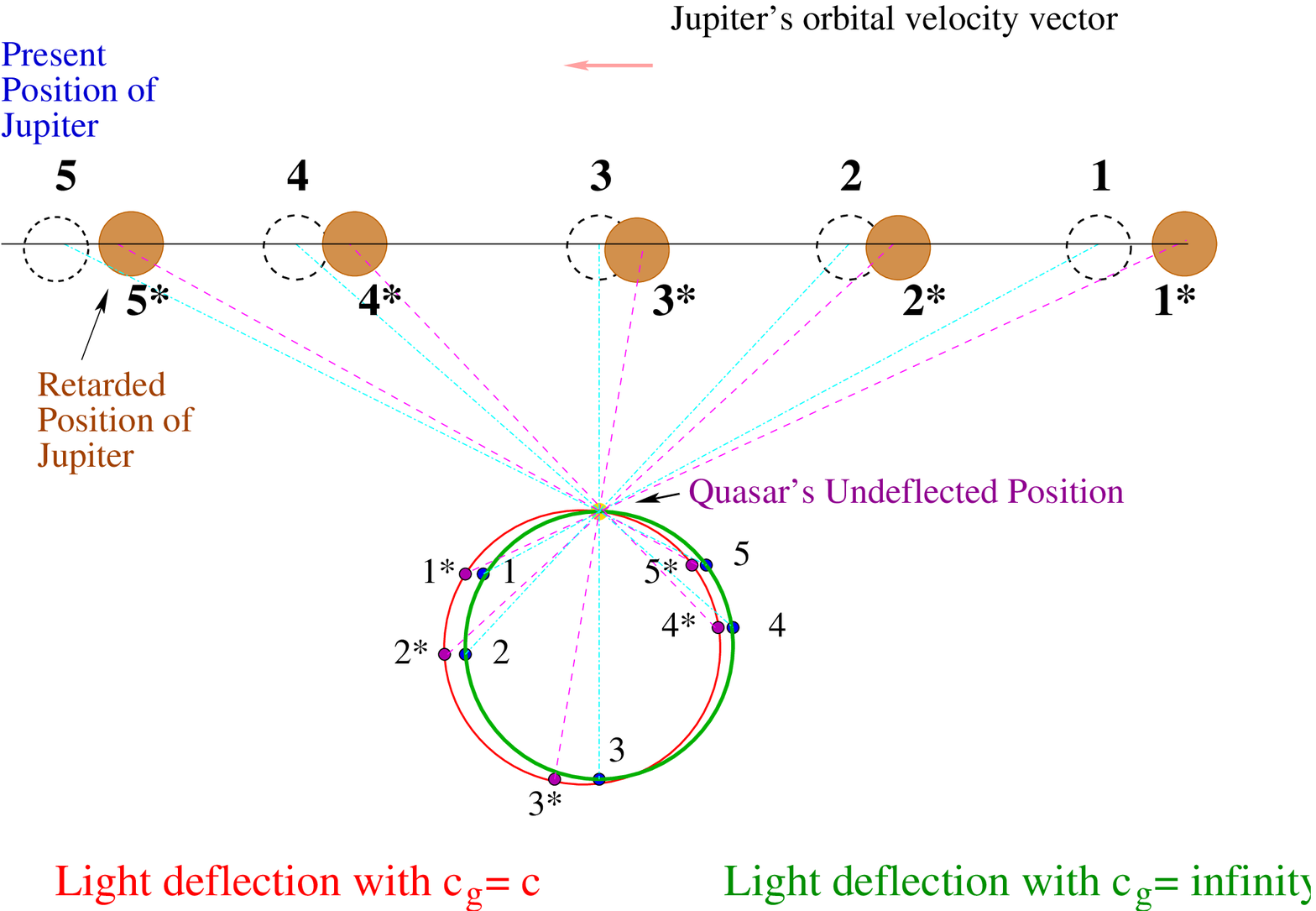}
\newpage
\caption[Deflection of Light by Jupiter]{While Jupiter moves in the sky from right to left the apparent position of the quasar J0842+1835 moves counterclockwise and completes a loop in the sky. 
If gravity propagates instantaneously, the apparent position of the quasar will move along the green circle 1-2-3-4-5 corresponding to the present positions 1, 2, 3, 4, 5 of Jupiter (white circles). The red curve 1*-2*-3*-4*-5* is our theoretical 
prediction for the light deflection of the quasar based on general relativity and the retarded positions of Jupiter (brown circles). Measuring the size and shape of the observed loop (especially in the neighborhood of the point 3*) allows us to evaluate the difference between the speed of gravity and light and to answer 
the question whether the light is deflected by the present position of Jupiter ($\delta=-1$) or by its retarded position ($\delta=0$) due to the finite speed of gravity \citep{fk2003}.  \label{fig1}}\vspace{-3cm}
\end{figure}

We have used the experimental data of the gravitational VLBI experiment on September 8, 2002 \citep{fk2003} to distinguish between the three cases for the parameter $\delta=0,-1,-2$, and to measure the speed of gravity $c_g$. We found \citep{fk2003} that $\delta=- 0.02\pm 0.19$ and the speed of gravity $c_g=(1.06\pm 0.21)c$. Two cases $\delta=0$ (red loop)  and $\delta=-1$ (green loop) are shown in Fig \ref{fig1}. Our experiment convincingly demonstrates that only retarded solution $\delta=0$ is consistent with the observations proving that the speed of gravity $c_g$ is equal to the speed of light $c$ within the experimental errors.

\acknowledgments
\noindent
We are thankful to J. Harris for valuable comments. The Epply Foundation for Research Award 002672 and support of the University of Missouri-Columbia Research Council were essential for completing this work.

\newpage
\end{document}